\documentclass[11pt,a4paper]{article}
\usepackage{caption}
\usepackage{subcaption}
\usepackage{jinstpub}
\usepackage{xcolor}
\usepackage{gensymb}
\DeclareUnicodeCharacter{2212}{-}

\title{26.5~ps Time Resolution Using 50~µm Low Gain Avalanche Detectors Fabricated by Micron Semiconductor Ltd.}

                                \author[a]{R.~Moriya,}
                                \author[a]{R.~Bates,}
                                \author[c]{M.~Bullough}
                                \author[a]{N.~Cooke,}
                                \author[a]{A.~Docheva,}
                                \author[a]{L.~Lombigit}
                                \author[a]{D.~Maneuski,}
                                \author[a]{R.~McFeely,}
                                \author[b]{N.~Moffat,}

                                \affiliation[a]{SUPA School of Physics and Astronomy, \\
                                           University of Glasgow\\
                                           University Avenue,Glasgow,Scotland}
                                \affiliation[b]{Centro Nacional de Microelectrónica, IMB-CNM-CSIC, Barcelona, Spain}
                                \affiliation[c]{Micron Semiconductor Ltd.}
                                \emailAdd{Richard.Bates@glasgow.ac.uk}
                                           \abstract{Low Gain Avalanche Detectors (LGADs) are silicon semiconductor sensors with an implanted thin p-doped multiplication layer that is designed to provide low gain. Most importantly, LGADs are specifically engineered to provide excellent spatial and temporal resolution simultaneously. The technology shows promising prospects of fulfilling the 4D tracking requirements of future high energy physics experiments.\newline \indent Micron Semiconductor Ltd. has fabricated LGADs with an active thickness of $50$~µm. The electrical and timing performance has been measured and compared with devices fabricated at IMB-CNM for reference. \newline \indent $50$~µm thin LGADs by Micron Semiconductor Ltd. were measured to have a timing resolution in the region of $30$~ps using a dedicated setup involving minimum ionizing particles produced by Sr-90. Specifically, the best timing resolution of $26.5$~ps was measured at a bias voltage of $200$~V at $-30$~$\degree$C.}
                                           \keywords{silicon detector, LGAD, APD, ps timing}
                                           \begin{document}
                                           \maketitle
\flushbottom

\section{Introduction}
In applications involving track reconstruction of charged particles in high energy physics (HEP) experiments, semiconductor detectors are widely used for the task of analysing the charge deposition in multiple layers of detectors. Traditional silicon tracking detectors provide a high spatial resolution in the order of $10$~µm derived from decades of intensive R\&D efforts and current HEP collider experiments such as the Large Hadron Collider (LHC) have low enough event densities to allow the reconstruction of every event through this technology. However, this outlook is expected to change with the upcoming High Luminosity (HL) LHC. The rate of events is expected to increase by approximately a factor of 5 compared to the luminosity in the LHC Run-2~\cite{rossi2015introduction}; this will result in a high event pile-up that would pose a challenge of separating individual tracks for reconstruction. However, the inclusion of precise timing information in the order of $10$~ps could solve this problem by discriminating compatible hits based on timing information in what is known as 4D tracking\cite{sadrozinski20174d}. Several candidate technologies have been proposed to fulfil this task in addition to other requirements such as but not limited to high radiation tolerance and costs.

One candidate technology is 3D silicon sensors which have a detector design with an electric field parallel to the detector surface, unlike traditional planar silicon sensors. This novel architecture was first introduced by Parker, et. al.~\cite{parker19973d} in 1997 and introduces advantages such as short drift time, radiation tolerance, and low depletion voltage\cite{CARTIGLIA2022167228,liu2020electrical,mendicino20193d}. Recent innovations in 3D silicon detectors have designs with small pixel pitches of $25$~µm $\times$ $100$~µm and $50$~µm $\times$ $50$~µm that provide excellent spatial resolution and temporal resolution\cite{dalla2016development}. Particularly, $50$~µm $\times$ $50$~µm devices were experimentally demonstrated to provide a peak time resolution in the order of $30$~ps\cite{kramberger2019timing}. However, the technology comes with several limitations. The manufacturing process of 3D silicon sensors is complicated\cite{dalla2022progress}. Furthermore, small pixel sizes needed for good time resolution increase power consumption due to the density of the electronic channels and the lack of gain in the technology will require appropriate circuitry for amplification that further increases power consumption\cite{CARTIGLIA2022167228}. 

Another candidate technology is complementary metal-oxide-semiconductor (CMOS), which has been available for decades and is actively being developed for HEP applications. Decades of R\&D have made CMOS-based detector systems easy and cheap to manufacture and sensor can be integrated into a single chip for easier assembly. Additionally, radiation damage in CMOS is well understood due to decades of R\&D and therefore the system can be optimized to be radiation tolerant. CMOS technology can be employed for 4D tracking by optimizing for spatial and temporal resolution. For example, FASTPIX is a monolithic CMOS sensor with pixel pitches tested from 8.66 um to 20 um and its design is optimized for timing through the application of a doping implant in the sensor to shape the electric field uniformly to improve the drift path of charges\cite{kugathasan2020monolithic,braach2022performance}. Furthermore, in contrast to traditional pixel sensors, FASTPIX uses a hexagonal sensor grid that shortens drift time and reduces charge sharing at the edges\cite{kugathasan2020monolithic,braach2022performance}. The demonstrator device has been experimentally measured to have a good timing resolution of 120 – 180 ps\cite{braach2022performance}. While CMOS technology shows promise, the current limitation is its trailing time resolution compared to 3D silicon sensors and the associated high power consumption due to small pixel sizes.

\begin{figure}
    \centering
    \includegraphics[width=1\textwidth]{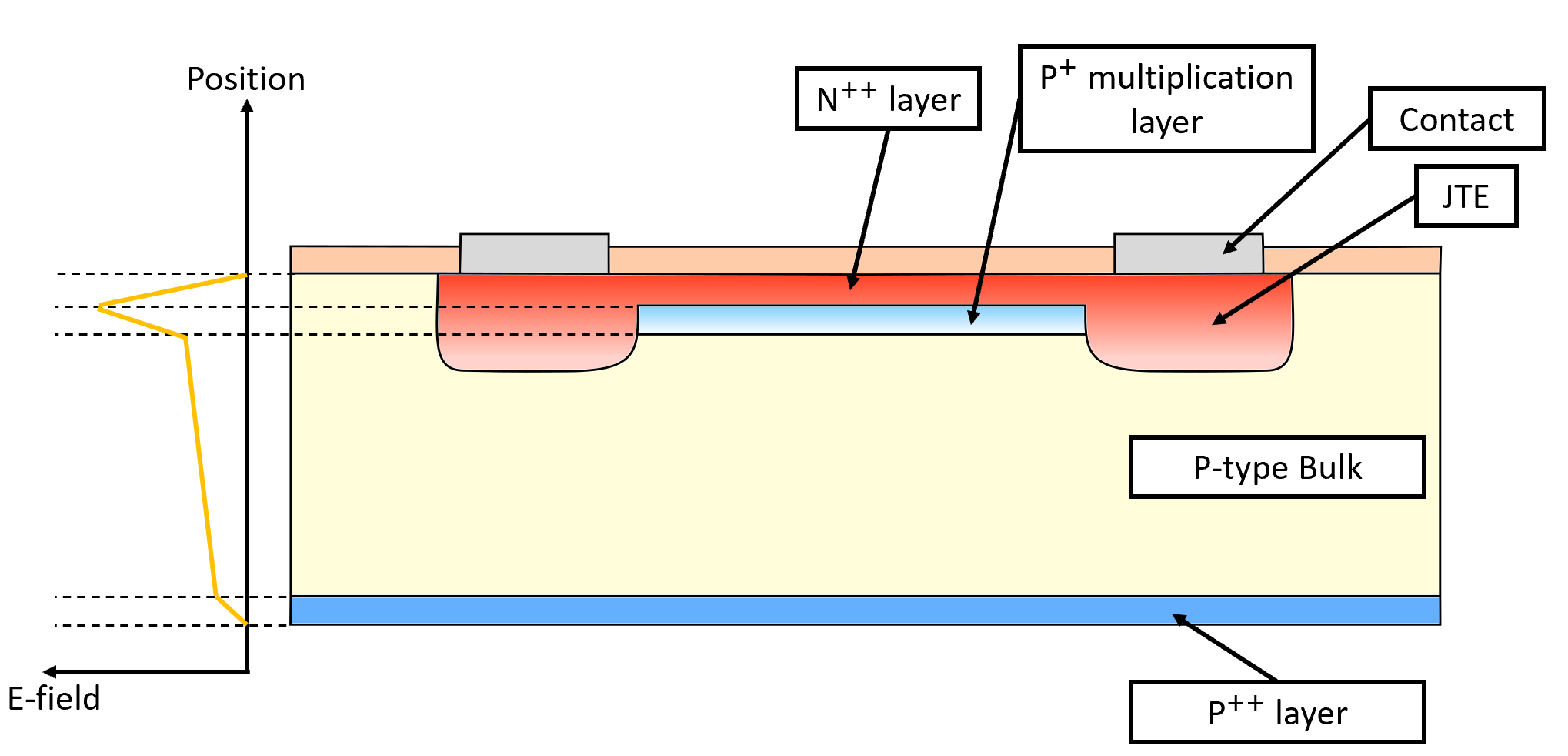}
    \caption{Schematic representation of LGAD (not to scale). The n++ and p+ layers are typically about $1 ‐ 3$~µm 
thick and the established electric field profile is depicted on the left.}
    \label{fig:schematic}
\end{figure}

Finally, another attractive option developed in the last decade is Low Gain Avalanche Detectors (LGADs) which is the paper’s main focus of study. LGADs are silicon pixel detectors optimised for timing performance\cite{moffat2018low} and a schematic is shown in figure~\ref{fig:schematic}. Particularly, LGADs have a low internal gain of about $10 − 20$ that is achieved by modifying the doping profile of a standard p-n junction\cite{sadrozinski20174d}. LGADs have excellent temporal and spatial resolution that are designed to fulfil the requirements of 4D tracking. Optimising the gain and thickness parameters allow a sensor with a fast charge collection time that could improve the time resolution in the order of $10$~ps\cite{sadrozinski20174d}. Excellent spatial resolution can be achieved through innovative designs such as AC-LGADs that utilise charge sharing between multiple electrodes and therefore surpasses the spatial resolution limitation of single pad detectors\cite{d2022measurements,apresyan2020measurements}. In the context of radiation tolerance, LGADs can fulfil this requirement by utilising carbon as a dopant for the gain layer and through the use of a narrower gain layer\cite{ferrero2019radiation}. Last but not least, an internal gain layer reduces the need for amplifier electronics and hence reduces power consumption for the operation of HL-LHC\cite{CARTIGLIA2022167228}. 

LGADs are an excellent candidate for 4D tracking and to understand their performance, LGADs were fabricated by Micron Semiconductor Ltd. The first timing measurement of LGADs by the vendor will be presented and compared to devices manufactured by the Institut de Microelectrònica de Barcelona (IMB-CNM) for reference. Moreover, the study will seek to understand how different physical parameters affect devices' timing resolution. The study will first focus on LGAD samples manufactured by both vendors and uncertainty factors affecting timing resolution will be elucidated in section~\ref{materials-methods}. Furthermore, the experimental setup for measuring the Current-Voltage (IV), Capacitance-Voltage (CV), gain, and timing resolution will be explored. Section~\ref{results-discussion} will present the experimental results obtained from the setup described in section~\ref{materials-methods} and subsequently interpreted to put into context. Finally, section~\ref{conclusion} will present potential issues encountered during the experiment and the outlook of the work.
\section{Materials and Methods}\label{materials-methods}
\subsection{Detector Design}
LGADs have been proposed as a new detector technology for use within a system that requires excellent time resolution. LGADs are based on the traditional Avalanche Photodiode (APD), however, the doping profiles are modified in such a way as to create the optimal electric field and hence low gain required for these detectors. As the gain of the LGAD is low, the device can be operated in avalanche mode and within the linear region with a large bias voltage working range and a gain that is loosely dependent on the bias voltage. The internal gain increases the sensor noise, but since electronic noise dominates the total noise, this contributes to an overall increase in the detector system’s signal-to-noise ratio.

The effective timing resolution of LGADs depends on four independent factors summed together: time walk, Landau noise, jitter, and time-to-digital converter (TDC). That is,
\begin{equation} \label{eq:1}
   \sigma^2_{Total} = \sigma^2_{Time Walk} + \sigma^2_{Landau Noise} + \sigma^2_{Distortion} + \sigma^2_{Jitter} + \sigma^2_{TDC}.
\end{equation}
The first term, the Time Walk, is defined as the uncertainty generated due to signals with different amplitudes crossing the same threshold at different times. The second term, the Landau noise, is the result of the variation of the charge deposited in each event that leads to the irregularity of the signal shape. Next, the distortion is the uncertainty caused by the non-uniform weighting field and variations in the drift velocity of the electron; these factors cause the signal's shape to change. Jitter is the uncertainty caused by the electronic noise that causes the threshold to be reached at different times. Finally, the time-digital converter (TDC) uncertainty is due to the effect of digitising the signal during readout.

The fabrication process of LGADs, therefore, involves fine-tuning the parameters such as thickness, gain and doping profile to minimise all the uncertainties described in equation~\ref{eq:1}. The time walk and jitter are minimised by maximising the slew rate $\frac{S}{t_{rise}}$ where $S$ is the amplitude and $t_{rise}$ is the rise time. Furthermore, jitter can be further minimised by increasing the gain of the device. The distortion can be minimised by operating the device at a high bias voltage to saturate the drift velocity and make use of a parallel plate capacitor geometry of the sensor to make the weighting field uniform. According to simulations, the Landau noise can be minimised by reducing the thickness of the sensor. Finally, the uncertainty due to the TDC can be neglected, as it does not make a significant contribution to the uncertainty and is not due to the sensor itself.

Based on these parameters in simulations, LGADs with the best timing resolution are achieved through thin devices and high gain. However, it must be emphasised that devices that are too thin will increase the capacitance of the devices, which have shown to worsen the time resolution. Optimising this balance through simulations has shown that devices with a thickness of $50$~µm with a gain in the magnitude of $10 - 20$ provide the best time resolution. Therefore, LGAD wafers fulfilling these technical requirements were fabricated by Micron Semiconductor Ltd and subsequently characterised at the University of Glasgow.

\subsection{LGAD Samples}
Timing studies were performed on the $50$~µm LGAD wafer fabricated from Micron Semiconductor Ltd. The wafer 3331-19 contains devices with varying pixel sizes, doping concentrations, and junction termination extension (JTE) width. For devices 1 - 7 in the wafer, the devices are single-pixel devices with pixel sizes varying from $0.22$~mm to $2$~mm and JTE widths ranging from $10$~µm to $50$~µm. Device 7 from the northern section of the wafer with a pixel size of $0.5$~mm and JTE width of $10$~µm was tested as part of the timing studies. As part of the timing measurement experiment, another device from the same region was used to ensure that the devices have similar geometry and characteristic parameters.

It is vital to understand what parameters affect timing resolution and make sure performance is on-par with other manufacturers of LGADs. Therefore, devices manufactured by CNM Barcelona were also measured. Particularly, $50$~µm LGAD devices W7-LGB-52 and W7-LGB-31 from run 9088 were studied. The two devices are single-pad LGADs with an active area of $3.3 \times 3.3$~$\textrm{mm}^2$.

\subsection{IV and CV: Setup}
The characterisation of the IV and CV curves of LGADs was done within the Glasgow Laboratory for Advanced Detector Development using a dedicated probe station setup. 

\begin{figure}[h]
    \centerline{\includegraphics[width=0.5\textwidth]{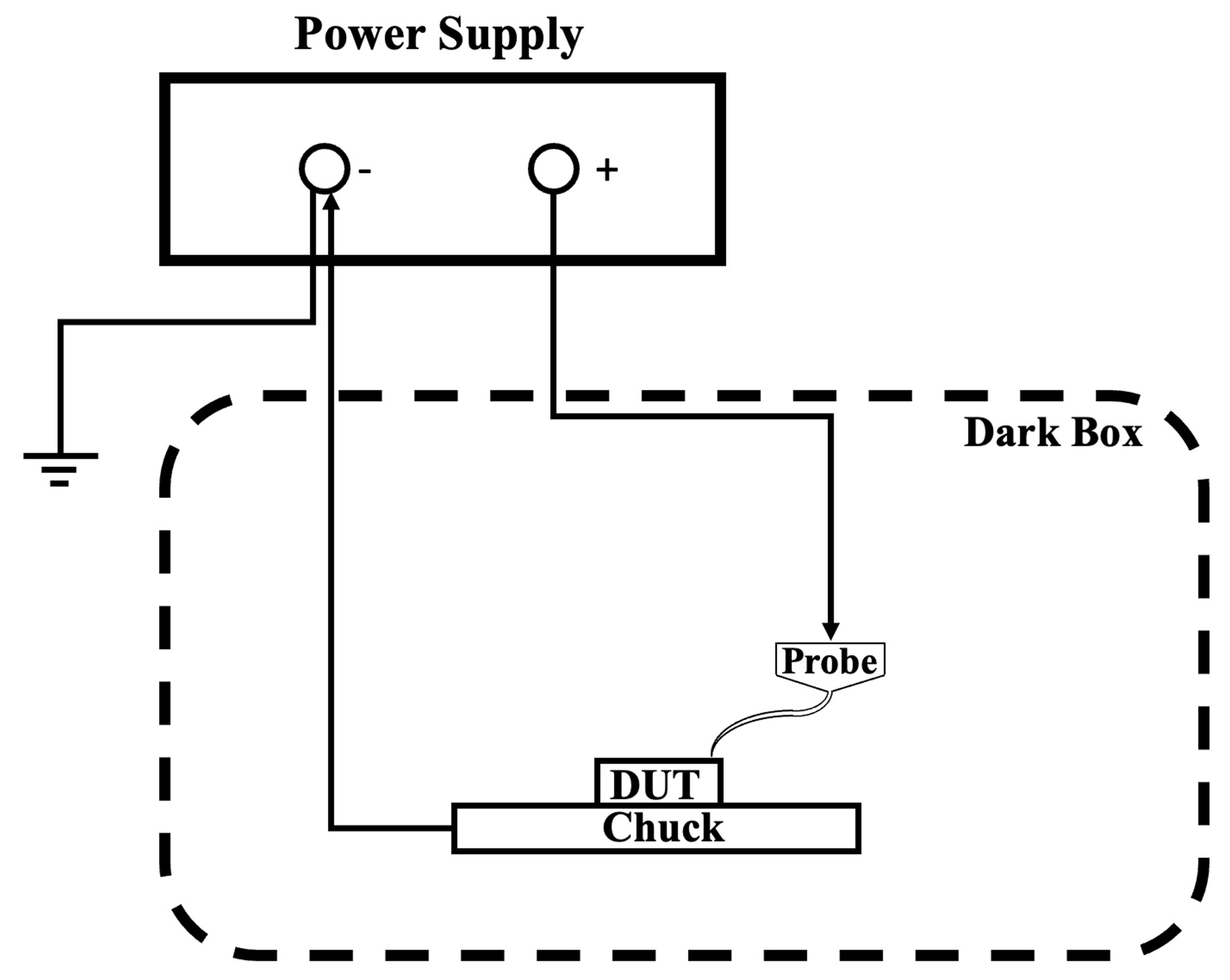}}
    \caption{\label{IV-setup} The setup used to take the IV measurements of the LGAD devices. The bias voltage is provided by the Keithley-273 power supply that is connected to the device and the chuck to be grounded. The device and the chuck are placed within the dark box to shield it from interference. }
\end{figure}

The main setup for characterising the IV curve consists of a power supply, a dark box, the probe, and the chuck. The setup is depicted in figure~\ref{IV-setup}. In this setup, devices are mounted on top of a chuck that is held in place by a vacuum pump. A probe composed of a tungsten filament of 20 microns is then lowered into contact with the device and a Keithley-273 DC power supply is connected to the probe through the positive terminal. For current to flow, the chuck is connected to the power supply at the negative terminal and is grounded. Finally, the entire setup apart from the power supply is placed inside the dark box from Wentworth to shield it from external interferences such as light and vibrations. During the IV measurements, the setup is connected to a computer and data is displayed through a dedicated LabView user interface.

The CV measurement setup is more complex than the IV setup. However, the two share similarities. Specifically, the CV setup uses the same probe station as the IV setup: the devices are mounted on the chuck positioned inside the dark box and are measured using the same probe. However, the CV circuit uses an LCR meter (Agilent-HP4284) in parallel mode for capacitance measurement and the power supply (Keithley-273) provides the bias voltage.

\begin{figure}[h]
    \centerline{\includegraphics[width=1\textwidth]{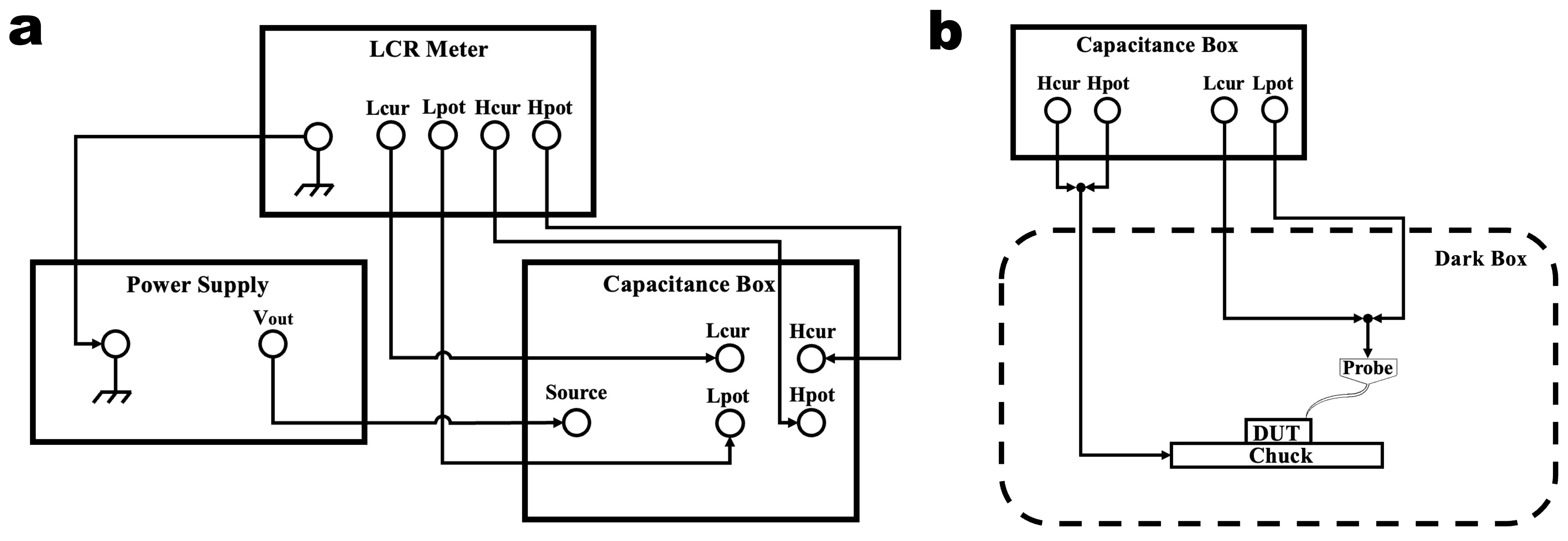}}
    \caption{\label{CV-setup} The setup used for taking CV measurements of the LGAD devices. (\textbf{a}) The front side of the panel contains the LCR meter and the power supply, which is used to measure the capacitance and provide the bias voltage respectively. (\textbf{b}) The back side of the panel is connected to the device and the chuck. Similarly to the IV setup, the measurements take place in the dark box to prevent any external interference. }
\end{figure}

\subsection{Gain: TCT Setup}\label{TCT_setup}
The transient current technique (TCT) is a versatile tool used for several decades to study semiconductor devices\cite{eremin1996development}. In this technique, laser light is used to create a signal in the semiconductor sensor and result in the generation of e-h pairs that is subsequently read out by the setup.

\begin{figure}[h]
    \centerline{\includegraphics[width=1\textwidth]{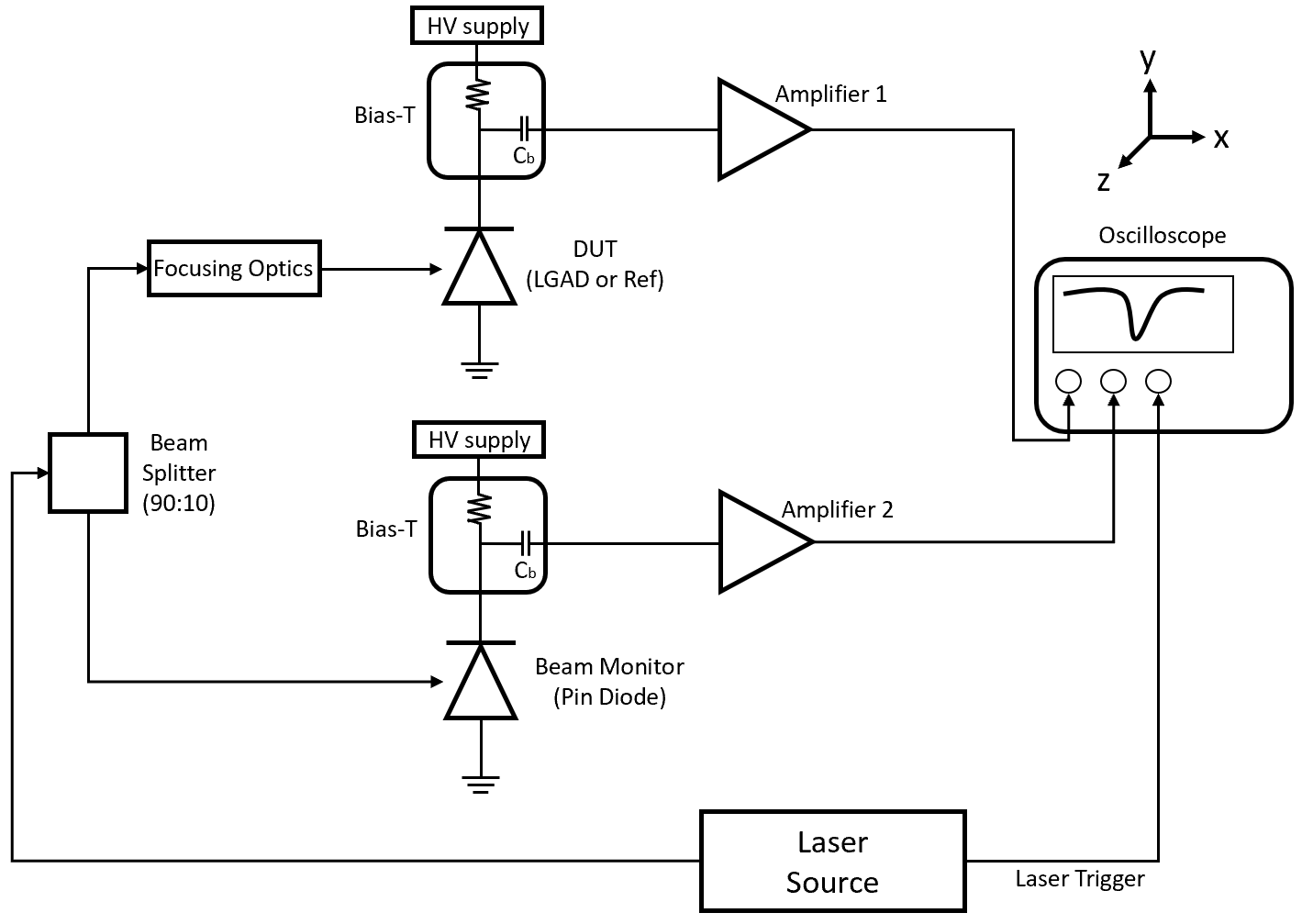}}
    \caption{\label{TCT_setup} A schematic of the TCT setup for measuring the gain of the devices. The laser output is split and fed to the two legs of the DUT and beam monitor at a ratio of $90:10$ respectively. The resulting signal from both sensors is then read and saved by the oscilloscope.}
\end{figure}

The setup is shown in figure~\ref{TCT_setup} and is supplied by Particulars. The experiment uses a diode laser ($660$~nm) that acts as a source for charged particles and has two outputs. The first output is the trigger signal that is fed to the oscilloscope to ensure the laser signal generated matches the signal generated from the device under test (DUT). The second output is split using a 90:10 beam splitter to be fed on two detectors. The two legs use a similar setup. A reverse bias voltage is applied on the detector using a high voltage (HV) power supply and an RF amplifier with a bandwidth of $0.01 - 2000$~MHz is used to improve the signal-to-noise ratio. Between the detector and the amplifier, a component known as bias-T is used. This serves to decouple the power from the HV power supply (DC) to the amplifier signal input (AC) to eliminate unwanted signals from appearing in the measurement. The main difference between the two legs is the inclusion of the focusing optics in the DUT leg that acts to focus the laser and the purpose of the two legs. The beam monitor leg includes a PIN diode used to monitor the output of the beam and benchmarked with previous measurements to quantify the laser fluctuation. This will ensure that the fluctuation in charge collected from a measurement taken in previous days remains the same. Furthermore, the PIN diode receives a weaker laser signal from the beam splitter to avoid saturation of the PIN diode. The actual measurements of the signal are done on the DUT leg. Gain can be obtained by running two different sets of measurements. The first measurement is to use a standard PIN diode with similar geometry and parameters as the DUT. The second measurement will use LGAD as the DUT. The gain can then be obtained by normalising the measurement's signal height using the beam monitor data and then subsequently comparing the two measurements. Particularly, the lack of gain in the PIN diode will be shorter than that of LGAD and hence making gain measurements possible. Last but not least, a Peltier cooling module was used in the setup (omitted in figure~\ref{TCT_setup}) to allow temperature control of the device down to $-30$~$\degree$C. The cool side is in direct contact with the aluminium support mounting the DUT and the hot side is connected to a water-based chiller.

\subsection{Timing Resolution: ${}^{90}\textrm{Sr}$ Setup and Analysis Procedure}
The section here will outline the setup used to perform time resolution measurements and the analysis procedure used to extract the time resolution. The setup was placed inside a climate chamber to control the temperature and dry air was constantly pumped to avoid issues resulting from humidity. The analysis procedure was done through a dedicated script written in the Python programming language.

\begin{figure}[h]
    \centerline{\includegraphics[width=1\textwidth]{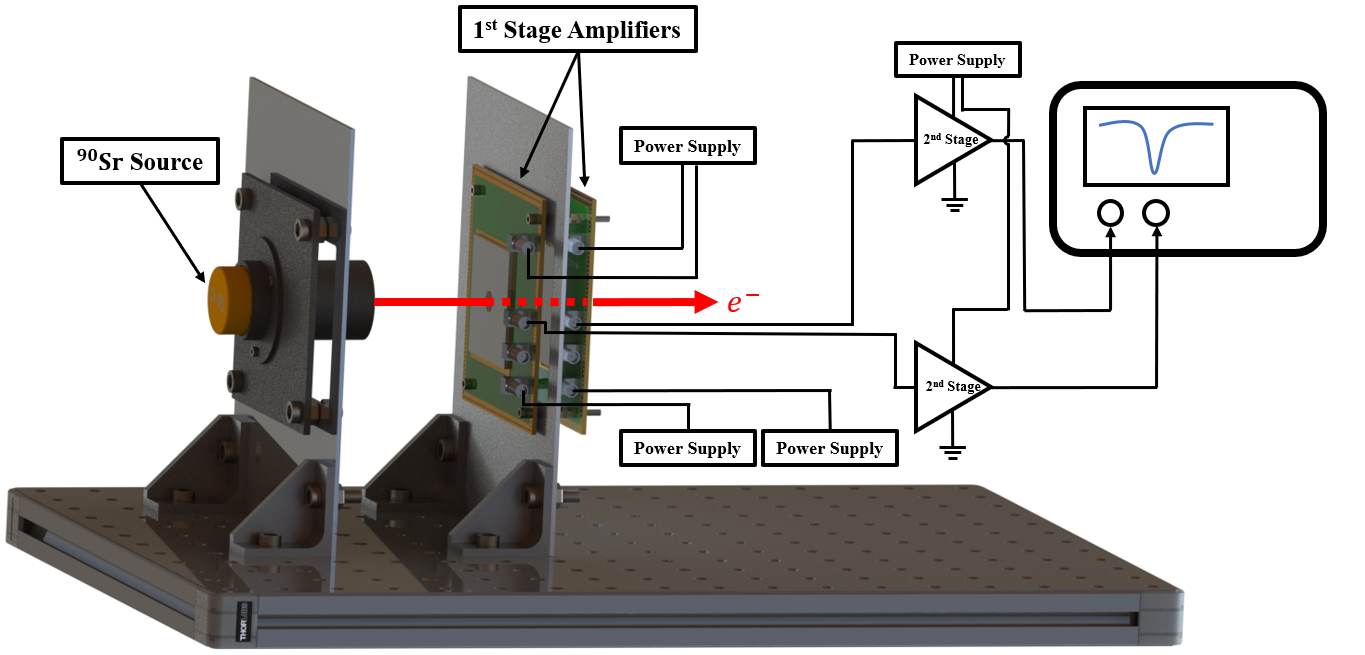}}
    \caption{\label{LGAD_setup} An illustration of the setup to measure time resolution. A Sr-90 source generates a beam of $\beta$-particles that passes through both detectors and the signal registered by the devices mounted on the PCBs, where they serve as 1st stage amplifiers. The PCBs are connected to 2nd stage amplifiers and then the signals are displayed by the oscilloscope. The entire setup is placed on top of a breadboard purchased at Thorlabs, Inc.}
\end{figure}

The setup is illustrated in figure~\ref{LGAD_setup}. Two planes of aligned LGADs are used to take coincidence measurements and $\beta$-particles from Sr-$90$ are used to generate signals in the detectors. A source holder was designed and printed using a 3D printer. Each LGAD are mounted on a dedicated Printed Circuit Board (PCB) that acts as an inverting trans-impedance amplifier and as a readout for the signals generated by the detectors. To ensure alignment with the source, screws were used in each corner to fasten the PCBs. The LGADs are mounted on top of a circular aperture to ensure that $\beta$-particles from the source can pass through the first detector and generate a signal in the second detector. Furthermore, aluminium shielding was placed on both sides of the PCB to protect the mounted devices from mechanical damage. Holes were drilled on both sides of the aluminium lid and aluminium tape with a thickness of $50$~µm was placed on the drilled holes to ensure particles from the source can hit the detector. Signals from the readout are then passed on into the 2nd stage amplifier, which is based on a discrete integrated Gallium Phosphate broadband amplifier. The amplified signal is then fed to an oscilloscope. The entire setup is placed inside a climate chamber to allow flexible temperature control. The oscilloscope used for the experiment is the 4 GHz - 8 bit vertical resolution MSO9404A Mixed Signal Oscilloscope by Keysight. This oscilloscope has a sampling rate of 20 GS/s and was found to provide an effective time discretisation of $3.13$~ps. The oscilloscope setting is set to trigger over threshold in each channel and the AND qualifier is applied. This is to ensure that the generated signals in both channels coincide. Finally, the waveform of the events was saved for offline analysis.

Constant Fraction Discrimination (CFD) was used as the analysis procedure of choice to yield the timing resolution from the data. In this procedure, the time of arrival of a particle incident on the detector is defined to be at a threshold on the rising edge of the signal that is a certain fraction of the signal's maximum height\cite{sadrozinski20174d}. Particularly, the time of arrival at thresholds from 10\% to 90\% at an increment of 10\% were extracted for both signals. The time difference can then be defined, where it is the difference between the time of arrival extracted from the front and rear detectors.  Using the same time difference thresholds, the time difference values were extracted for each set of events and subsequently binned into a histogram. The number of bins in the histogram could be manually set to some reasonable value. However, the optimal bin size was dynamically selected based on a method used to minimise the mean integrated squared error associated with the statistics. The resulting histogram is a Gaussian distribution and its sigma is what is known as the time resolution of the detectors. This calculated value is the system's average timing resolution and to calculate the individual devices' timing resolution, an important assumption was made in the analysis: it is assumed that the detectors have similar time resolution. This comes from the fact that identical detectors from the same region of the wafer were selected. As there are two planes of detectors, the calculated timing resolution was divided by a factor of $\sqrt{2}$. Furthermore, as part of the analysis, different combinations of constant fractions from both detectors were used. For example, the front detector's 50\% to the rear detector's 30\%. As such, a total of 81 timing resolution values were calculated and the best timing resolution from the combinations was extracted from each measurement run. Finally, the uncertainty from the measurements was extracted by the following procedure: the measurement was repeated 5 times for a certain bias voltage value with the entire setup being turned off for an hour between each measurement. The measurement's standard deviation was then taken to obtain the uncertainty. 
\section{Results and Discussion} \label{results-discussion}
\subsection{IVs and CVs}
\begin{figure}[h]
    \centerline{\includegraphics[width=1\textwidth]{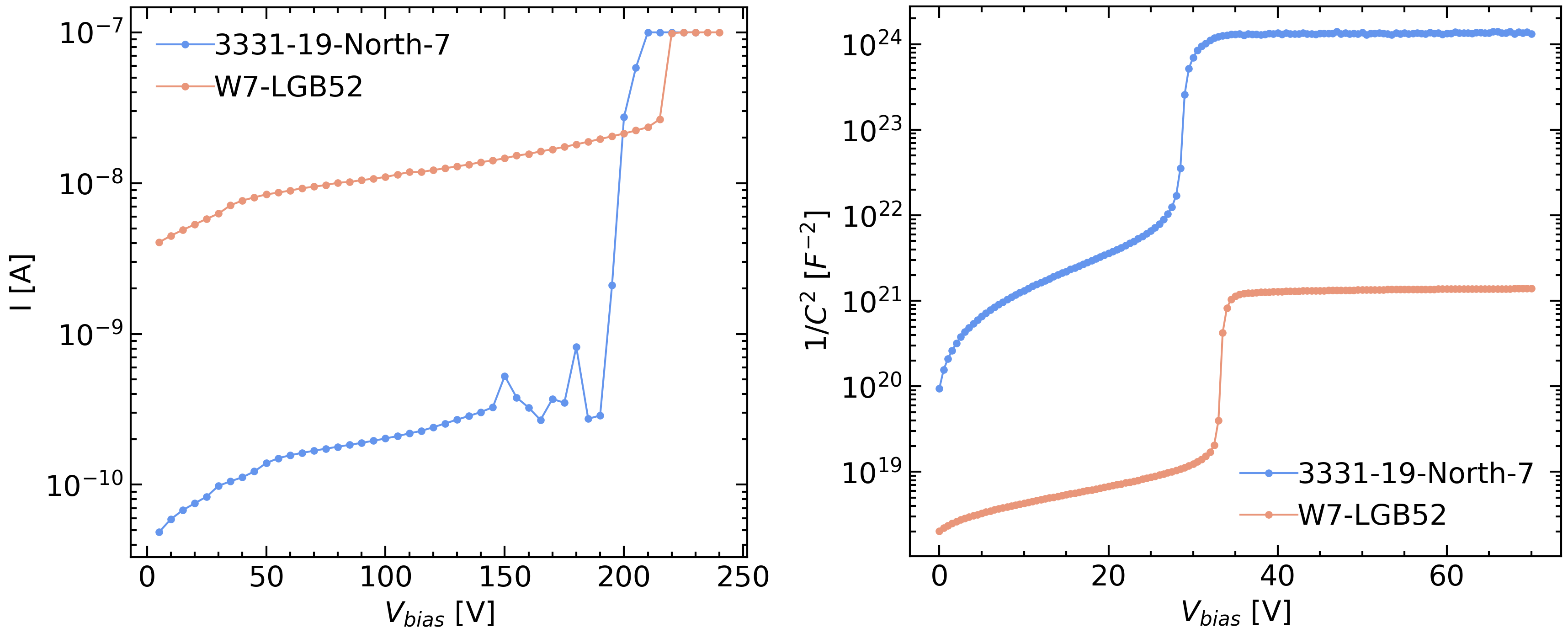}}
    \caption{\label{IV-CV} The IV (left) and CV (right) curves of W7-LGB-52 and 3331-19-North-7 devices. }
\end{figure}

Basic performance characterisation of LGADs were performed on both devices at room temperature. Figure~\ref{IV-CV} shows the Current-Voltage (IV) and Capacitance-Voltage (CV) measurements. 

Applying a bias voltage on the device initially results in the production of leakage current that increases with higher voltage. However, there is a threshold which results in a sudden jump in current and this characteristic voltage is known as the breakdown voltage. For the Micron device, this happens at approximately $195$~V while it happens at approximately $220$~V for the CNM device. Data further indicate that leakage current is higher for the CNM device than the Micron device.

The CV measurement similarly shows a jump in its capacitance before reaching a plateau. The CV curve has two transitions: the rapidly rising edge and the edge that rapidly plateaus. The former corresponds to the voltage which depletes the multiplication layer and the latter is the full depletion voltage of the sensor that happens a few volts after the former. The Micron device has its multiplication layer depleted at around $27$~V and fully depleted at around $30$~V. In contrast, the CNM device has its multiplication layer depleted at around $32$~V and full depletion at $35$~V. This shows that depletion voltage is higher for the CNM device compared to the Micron device.

\subsection{Gain}
\begin{figure}[h]
    \centerline{\includegraphics[width=1\textwidth]{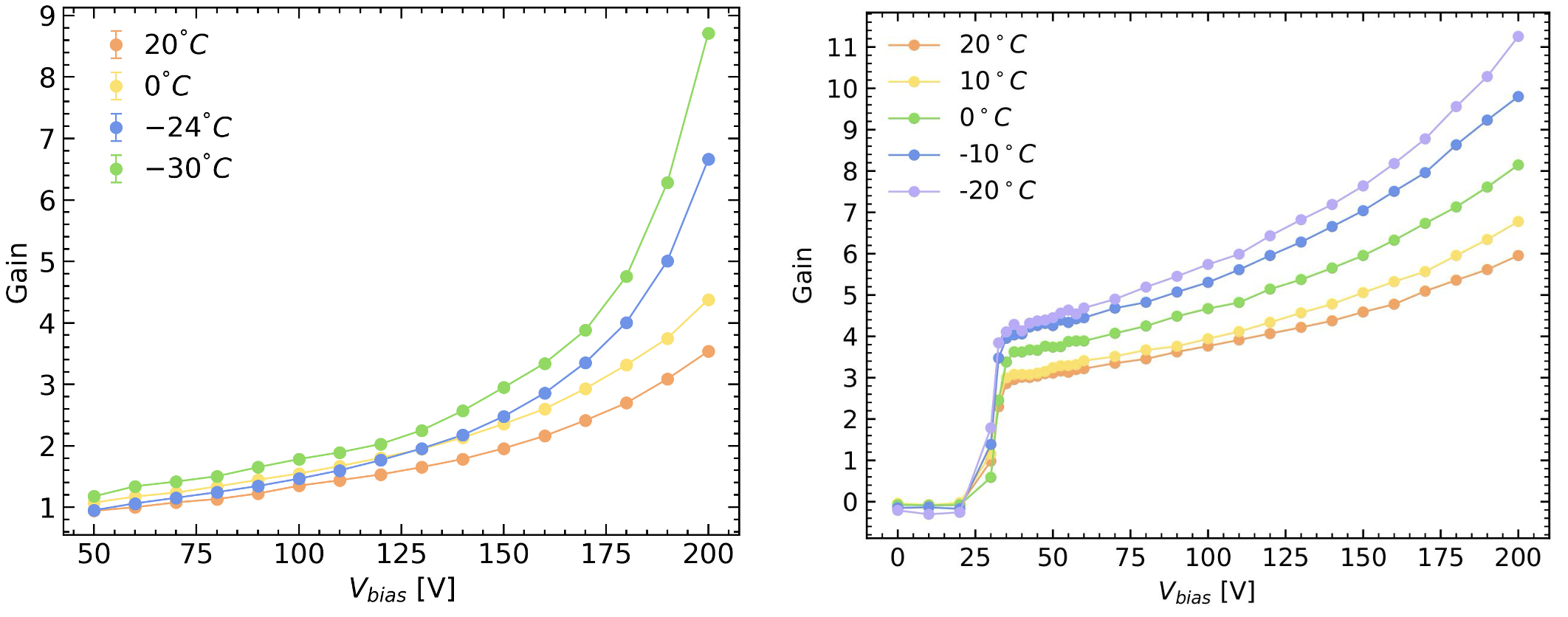}}
    \caption{\label{gain-voltage} The Gain as a function of detector bias voltage for W7-LGB-31 (left) and 3331-19-North-7  (right).}
\end{figure}

The gain of the devices was measured using the TCT setup described in section~\ref{TCT_setup}. The Gain as a function of bias voltage for various temperatures is presented in figure~\ref{gain-voltage} for both devices.

\begin{figure}[h]
    \centerline{\includegraphics[width=1\textwidth]{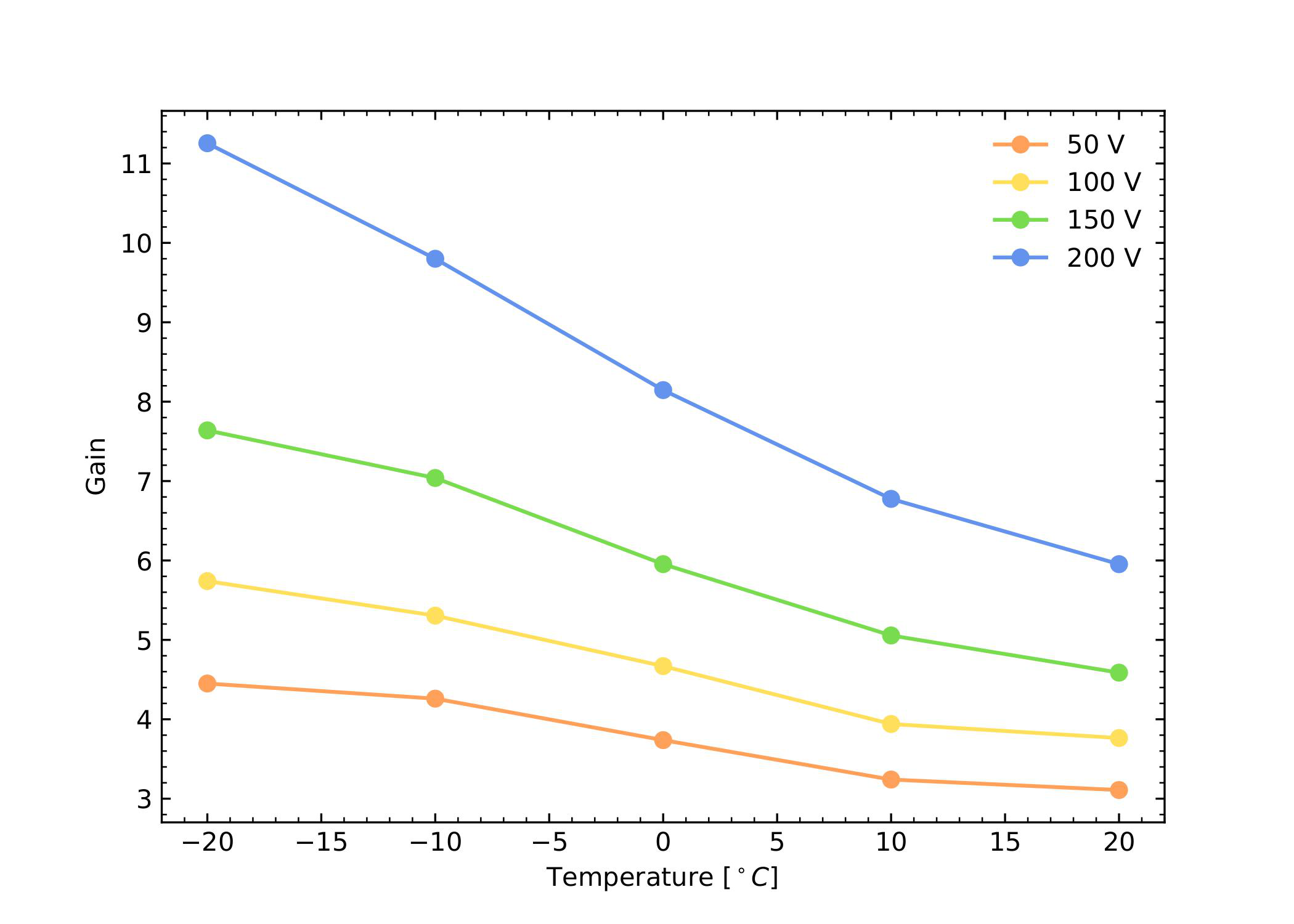}}
    \caption{\label{gain-vs-temp} The Gain as a function of temperature for 3331-19-North-7.}
\end{figure}

The gain for CNM devices starts at around 1 for all temperatures at $50$~V and remains approximately constant for the bias voltage range of $5$~V $-$ $130$~V. After around $130$~V, the gain increases at a rapid pace, especially for lower temperatures. As characteristic of LGADs, the gain is shown to be low and for this particular device, it is lower than 10. The behaviour in the gain for LGAD devices by Micron Semiconductor Ltd. was measured to be similar to CNM's LGADs, but the gain after depletion begins around 3 and 4 respectively for higher and lower temperatures. These data show that gain increases with increased bias voltage and with decreasing temperature. This is further supported by figure~\ref{gain-vs-temp}. Furthermore, it shows that gain increases slowly with decreasing temperature for lower bias voltages but increases more rapidly with higher bias voltages. 

\subsection{Timing Resolution}
\begin{figure}[h]
    \centerline{\includegraphics[width=1\textwidth]{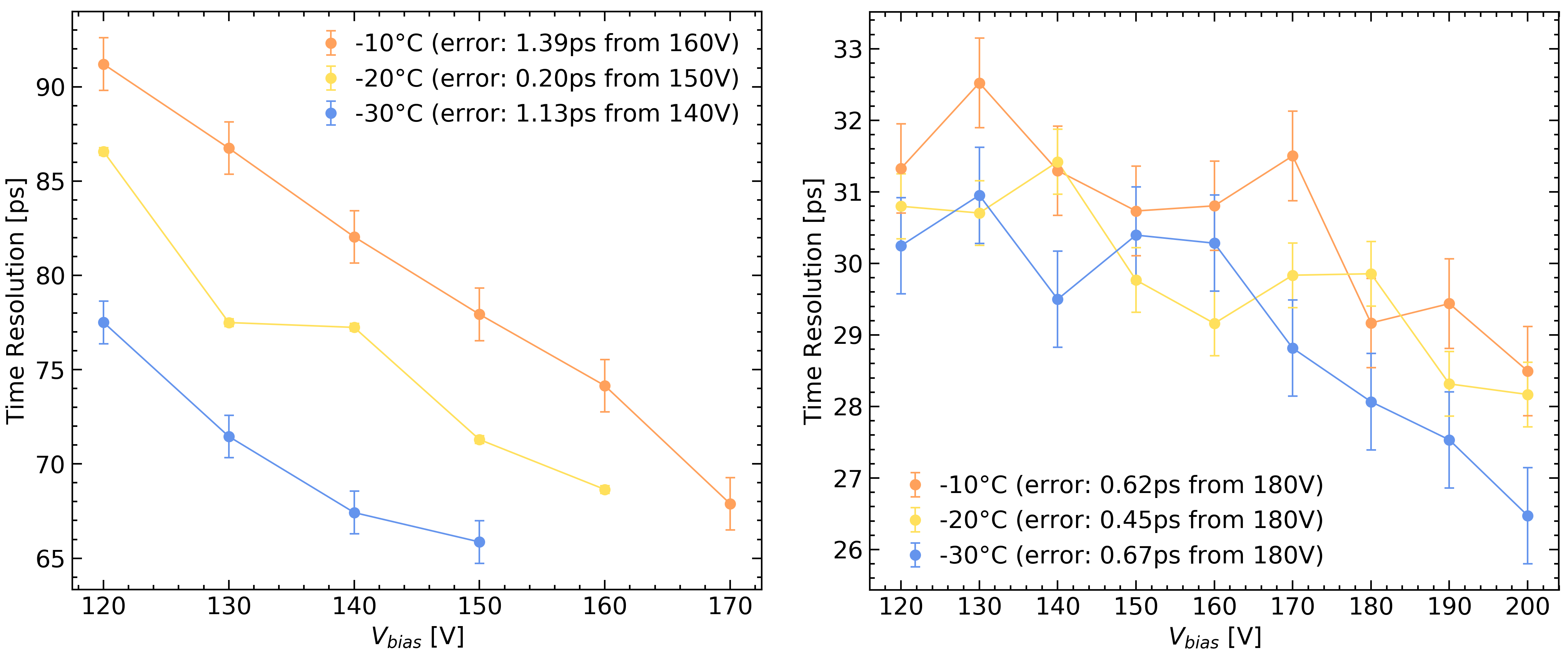}}
    \caption{\label{Time-resolution} The time resolution as a function of detector bias voltage for CNM (left) and Micron (right) devices at various temperatures.}
\end{figure}

The timing resolution for both CNM and Micron devices was measured for various temperatures with varying bias voltage. A total of $4000$ events were used for each data point as part of the analysis. Furthermore, the best timing resolution was taken out of all the threshold combinations for each bias voltage. The resulting data is shown in figure~\ref{Time-resolution}. Specifically, both devices were measured
at temperatures of $-10$~$\degree$C, $-20$~$\degree$C, and $-30$~$\degree$C but Micron devices were measured at a bias voltage range of $120$~V $-$ $200$~V in contrast to CNM devices' $120$~V $-$ $170$~V range. 

For Micron devices, lower bias voltage results in the timing resolution in the lower $30$~ps and the higher bias voltage results in better timing in the upper $20$~ps. The best timing resolution of $26.5$~ps was achieved at the highest bias voltage of $200$~V and lowest temperature of $-30$~$\degree$C. Furthermore, lower temperatures result in better overall timing resolution. The results for the CNM devices show a similar but clearer trend of better timing resolution with higher bias voltage and lower temperature. The best timing resolution of mid-$60$~ps was achieved at $150$~V at a temperature of $-30$~$\degree$C and the lowest timing resolution of lower $90$~ps was achieved at $120$~V at a temperature of $-10$~$\degree$C. The error was found to be $\sim{1}$~ps for both devices. 

\subsection{Discussion}
A holistic observation of all the results reveals important factors that influence the timing resolution of LGADs. The time resolution of LGADs improves with higher bias voltage and lower temperatures. This is because figure~\ref{gain-voltage} indicates that gain increases for higher bias voltage and lower temperatures that result in the gain reaching closer to $10 - 20$.

Comparing the time resolution of both devices provides further insight into the time resolution of LGADs. Best timing performance using CFD was achieved for thresholds ranging from $20\% - 50\%$ for CNM devices while for Micron devices, it corresponded to a narrow range of around $20\% - 30\%$. One possible explanation of better timing performance at a lower threshold for both devices is due to the signal strength being similar at the base of the signal and hence yielding a more consistent time difference resulting in a narrower time difference histogram. It is observed that for the same bias voltage range, the time resolution of Micron devices appears to be better than CNM devices. One possible reason is due to CNM devices' larger pixel size of $3.3$~mm compared to Micron devices' $0.5$~mm which results in higher capacitance (as shown in figure~\ref{IV-CV}) and therefore worse timing performance. Another explanation is that despite the higher breakdown voltage of the CNM devices, the time resolution was only measured up to $170$~V compared to Micron devices' $200$~V. This was because of CNM devices' high leakage current that made higher bias voltage measurements difficult due to its contribution to smearing the signal. Overall, measurements on LGADs manufactured by Micron Semiconductor Ltd. indicate that the timing performance is on par with LGADs from other manufacturers. For example, similar time resolution measurements using pairs of LGADs was done on other devices manufactured by CNM that achieved time resolution in the upper $20$~ps and lower $30$~ps.
\section{Conclusion}\label{conclusion}
LGADs are promising technology for 4D tracking in future HEP experiments due to their simultaneous excellent spatial and temporal resolution. The best time resolution for LGADs can be achieved by optimising the geometry and gain of the devices. To fulfil these requirements, LGADs were produced by Micron Semiconductor Ltd. and their timing performance was characterised for the first time. In addition, available LGAD devices from CNM were measured in parallel to understand if the devices were on par with other manufacturers and to understand parameters affecting timing resolution. Basic IV and CV measurements were taken for the devices. Then, a dedicated TCT setup was used to measure the gain of the devices. An experiment involving Sr-90 was set up and the measurements were analysed using CFD to quantify the time resolution of the LGADs. Results indicate timing resolution is strongly dependent on bias voltage and temperature due to their influence on the gain of the devices. Importantly, the time resolution of LGAD devices from Micron Semiconductor Ltd. was measured to be between the lower $30$~ps and upper $20$~ps and hence the produced LGAD devices are in line with other manufacturers.

During the measurements, several issues were present. The IV and CV measurements in figure~\ref{IV-CV} were found to be satisfactory. A clear breakdown voltage was observed for both devices, with a slight kink present in the IV measurements for the Micron device. The results for the gain measurement in figure~\ref{gain-voltage} show that gain from $50$V $-$ $120$V is higher for $0 \degree$C compared to $-24 \degree$C. This indicates a miscalibration in the TCT setup that systematically affected the normalisation of the data points. Nevertheless, the temperature dependence of the gain is observed. There are several limitations present in the analysis. First is the analysis method itself. CFD is a relatively simple method using the rising edge of the signal to quantify the signal and calculate the time resolution. However, it may not yield the most accurate and better time resolution compared to an alternative but more complicated procedure known as multiple sampling that makes full use of the signal. Another issue is with the devices. The measurements for IV, CV, and gain for the CNM device W7-LGB-31 were omitted due to the inability to characterise the device due to high leakage current and made time resolution measurements for bias voltage closer to breakdown voltage difficult. Another limitation is the assumption made during the analysis. Specifically, the devices’ time resolution can be calculated by dividing but this only holds if the devices’ parameters and geometry are the same. In reality, the devices may not be an exact copy and therefore may introduce additional errors to the measured time resolution of the devices. Last but not least, the error was measured for only one data point for each temperature instead of all the points.

The performance results for LGADs manufactured by Micron Semiconductor Ltd. show promising prospects for fulfilling the 4D tracking requirements for future HEP experiments and other applications. Future work will focus on measuring timing performance after irradiation of the devices. This will ensure that timing performance is still up to standards after being exposed to the radiation-harsh environment of HEP experiments.

\bibliography{mybib} 
\bibliographystyle{ieeetr}
\end{document}